\documentclass{gm-mpla}

\begin{document}

%

\def\nocropmarks{\vskip5pt\phantom{cropmarks}}

\let\trimmarks\nocropmarks      

%

\markboth{H.~Gao \& B.-Q.~Ma} {Exotic Hadrons of Minimal
Pentaquark ($qqqq\bar{q}$) States}

%
\catchline{}{}{2313}{}{}
\setcounter{page}{2313}

\title{EXOTIC HADRONS OF MINIMAL PENTAQUARK \\
($qqqq\bar{q}$) STATES }

\author{\footnotesize HAIYAN GAO\footnote{Now at Department of Physics, Duke
University, Durham, NC 27708, USA. ~gao@tunl.duke.edu} }

\address{Laboratory for Nuclear Science and Department
of Physics, \\
Massachusetts Institute of Technology, Cambridge, Massachusetts
02139, USA}

\author{BO-QIANG MA\footnote{Now at Department of Physics, Peking University, Beijing 100871,
China. ~mabq@phy.pku.edu.cn}}

\address{CCAST (World Laboratory), P.O.~Box 8730, Beijing
100080, China\\ and \\
Institute of High Energy Physics, Academia Sinica, Beijing 100039,
China}

\maketitle

\pub{Received 18 October 1999}{Originally published in Mod. Phys.
Lett. A, Vol.~14, No.~33 (1999) 2313-2319}

\begin{abstract}
It is shown that the exotic non-$qqq$ hadrons of pentaquark
$qqqq\bar{q}$ states can be clearly distinguished from the
conventional $qqq$-baryon resonances or their hybrids if the
flavor of $\bar{q}$ is different from any of the other four
quarks. We suggest the physical process
$p(e,e'K^-)Z(uuud\bar{s})$, which can be investigated at the
Thomas Jefferson National Accelerator Facility (JLab), as an ideal
process to search for the existence or non-existence of the exotic
hadron of minimal pentaquark state $Z(uuud\bar{s})$.
\\
\hspace*{16pt} {The search for the existence of $Z(uudd\bar{s})$
is also discussed in the paper.\footnote{We added this sentence
after the original published form of this paper.}}

\end{abstract}


\vspace*{6pt}

\noindent The quark model\cite{QM} has been proved to be
remarkably successful in classifying the mesons and baryons as
composite systems of quark-antiquark ($q\bar{q}$) states for
mesons and three quark ($qqq$) states for baryons.\cite{PDG} Also,
it has been well known that the structure of hadrons should be
more complicated than the simple lowest valence quark states, and
there are additional sea quark-antiquark pairs and gluons inside
the hadrons probed by various deep inelastic processes. The
situation can be well illustrated by the structure of the nucleon
in terms of quark-gluon components as supported by a large number
of phenomenological evidence. The existence of the higher
quark-gluon components in the nucleon structure can be described
by perturbative quantum chromodynamics (pQCD) for the perturbative
aspect and perhaps by the baryon-meson fluctuation
configuration\cite{Bro96} for some of the non-perturbative
aspects. Therefore a conventional baryon should be composed of
various quark-gluon configurations in which the lowest is the
three quark ($qqq$) state. From another point of view, there have
been various theoretical and experimental investigations on the
possibility of the existence of exotic hadrons beyond the
conventional quark model spectroscopy, such as multiquark mesons
($qq\bar{q}\bar{q}$) and baryons ($qqqq\bar{q}$), dibaryons
($qqqqqq$), hybrid states ($q\bar{q}g$), and glueballs.\cite{PDG}
The present situation seems to be very promising for the existence
of the hybrid states and glueballs. Though much progress has been
made and many candidates for the other exotic hadrons have been
reported, the existence of the exotic hadrons of multiquark states
is still far from being clear, due to the difficulty to clearly
distinguish candidates from a large number of various possible
meson and baryon resonances.

In this letter we will focus our attention on the exotic hadrons
composed of five quarks, i.e., a new class of particles of minimal
pentaquark ($qqqq\bar{q}$) states. There have been several
candidates for the pentaquark states, such as
$\Lambda(1405)$,\cite{PDG,l1405} the possible existence of the
pentaquark baryons of hidden strangeness $X(2050)$ and
$X(2000)$,\cite{Lan99} and several possible narrow states which
might be pentaquark states.\cite{nbr} The $\Lambda(1405)$ has been
confirmed to exist,\cite{PDG} but it is still not clear whether it
is a three quark $uds$ state resonance or a baryon-meson
state.\cite{l1405} The existence of several other possible
pentaquark states still needs to be confirmed. Although these
possible particles may have properties different from conventional
$qqq$-baryon, it is difficult to identify them unambiguously from
possible baryon resonances or their hybrids. In the following we
will show that the pentaquark ($qqqq\bar{q}$) states with the
antiquark $\bar{q}$ having different flavor from any of the other
four quarks have the unique property of no place in the
spectroscopy of the $qqq$-baryon quark model, and can not be
misidentified with any of the baryon resonances or hybrids. Thus,
they might be helpful to clarify the present situation concerning
$\Lambda(1405)$ and to confirm the existence of the pentaquark
states if any evidence for such a hadron can be observed.

For simplicity, we will consider the pentaquark ($qqqq\bar{q}$)
states as composite systems of baryon-meson states, inspired by
our understanding of the non-perturbative sea quark-antiquark
pairs of the nucleon in terms of baryon-meson
fluctuations.\cite{Bro96} Although the baryon-meson states can not
fully accommodate all possible pentaquark states, it will be
sufficient as a first approximation to see whether it provides the
hint for a new class of particles. It has been pointed
out\cite{Bro96} that the lowest baryon-meson fluctuations of the
nucleon can provide a comprehensive understanding of a number of
empirical anomalies, such as the violation of the Gottfried sum
rule\cite{Kum97}, the large excess of charm quarks at large
Bjorken $x$,\cite{EMC82,Bro81} the strange quark-antiquark
distribution asymmetry,\cite{Bro96} and the violation of the
Ellis-Jaffe sum rule.\cite{SpinR} Thus any of the conventional
baryons in the quark model should have an important part of
baryon-meson components beyond the three valence quark
$qqq$-states. Such baryon-meson fluctuations are of
non-perturbative nature and can be handled by models at the
moment, but their existence can be verified and studied from
various experiments. There are also theoretical studies which show
that the $\Lambda(1405)$ may be dominated by baryon-meson terms in
the wavefunctions.\cite{l1405} Also it has been found that there
is increasing empirical support for an $\eta$-baryon octet of
$J^p=\frac{1}{2}^-$ states associated with the $S$-wave $\eta+N$
[$N(1535)$], $\eta+\Lambda$ [$\Lambda(1670)$], $\eta+\Sigma$
[$\Sigma(1750)$] threshold interactions.\cite{Tuan} These results
naturally inspire us to consider the possible existence of a new
class of particles in terms of baryon-meson states. Some of these
baryon-meson states can be served as the intrinsic quark-antiquark
sea components in the ordinary $qqq$-baryons and they mix with the
ordinary baryons, but we will show that some pentaquark
($qqqq\bar{q}$) states in terms of meson-baryon states can not
find their place in the conventional three quark ($qqq$) baryon
spectroscopy.

We notice that in case the antiquark has the same flavor (e.g.,
$q_4$) with any of the other four quarks in the pentaquark
$q_1q_2q_3q_4\bar{q}_4$ state, such pentaquark state may be easily
identified with the three quark state of $q_1q_2q_3$-baryon
resonance, from the above knowledge that any baryon of $q_1q_2q_3$
valence states may have higher quark components of
$q_1q_2q_3q_4\bar{q}_4$ states. We also notice that all of the
previous reported candidates of pentaquark states belong to such
$q_1q_2q_3q_4\bar{q}_4$ states with hidden flavor $q_4$, thus they
are difficult to be distinguished from possible $q_1q_2q_3$-baryon
resonances with exotic properties. A completely unambiguous way of
identifying a pentaquark $qqqq\bar{q}$ state is that the antiquark
$\bar{q}$ having flavor different from any of the other four
quarks. We find that such pentaquark states can not exist in the
quark model $qqq$-baryon spectroscopy. Thus, they have the unique
property of minimal configuration $qqqq\bar{q}$ to be
distinguished from the conventional baryons.

Such minimal pentaquark states composed of the light flavor up
($u$) and down ($d$) quarks can only exist in two possible states:
$uuuu\bar{d}$ and $dddd\bar{u}$, which can be considered as
composite systems of $\Delta^{++}(uuu)\pi^+(u\bar{d})$ and
$\Delta^{-}(ddd)\pi^-(d\bar{u})$. We find that it is not easy to
find good processes which can produce such particles conveniently.
There are more such minimal pentaquark states composed of three
flavor $u$, $d$, and $s$ (strange) quarks (but with only one
$\bar{s}$ or $s$ quark): $uuuu\bar{s}$, $uuud\bar{s}$,
$uudd\bar{s}$, $uddd\bar{s}$, $dddd\bar{s}$, $ddds\bar{u}$, and
$uuus\bar{d}$. When consider them in terms of baryon-meson states,
those in which the baryon is the nucleon should be:
$uuud\bar{s}=p(uud)K^+(u\bar{s})$,
$uddd\bar{s}=n(udd)K^0(d\bar{s})$, and
$uudd\bar{s}=n(udd)K^+(u\bar{s})=p(uud)K^0(d\bar{s})$. From which
we notice that the minimal pentaquark states which can be easily
measured might be $uuud\bar{s}=p(uud)K^+(u\bar{s})$ and
$uudd\bar{s}=n(udd)K^+(u\bar{s})$, which might be produced from
physical processes $p(e,e'K^-)Z(uuud\bar{s})$ and
$n(e,e'K^-)Z(uudd\bar{s})$ \footnote{Since there is no free
neutron target, deuterium targets should be used in this case.}
which can be measured at the Thomas Jefferson National Accelerator
Facility(JLab). As far as we know, there have been theoretical
studies on the properties of the pentaquark states in the MIT bag
model\cite{MIT} and in other frameworks.\cite{penta} The
pentaquark states we suggest to measure are the ones with
strangeness number $S=1$ (i.e., with an anti-strange quark inside
the hadron), a distinct quantum number that can not be possessed
by ordinary baryons in the conventional quark model spectroscopy.
There have been also suggestions for the possible existence of
other exotic multiquark states, such as $\bar{c}sq\bar{q}$ and
$\bar{c}sqqq$ with both strangeness and charm.\cite{li} However,
there is still no convincing experimental evidence for the
existence of the above mentioned minimal pentaquark states such as
$Z(uuud\bar{s})$.

There should be a number of physical processes that could in
principle produce the minimal pentaquark state $Z(uuud\bar{s})$,
such as $p(\gamma, K^-)Z(uuud\bar{s})$,
$p(p,p'K^-)Z(uuud\bar{s})$, and $p(\pi^+,K^0)Z(uuud\bar{s})$ {\it
et al.}. The $p(e,e'K^-)Z(uuud\bar{s})$ process at JLab is ideal
for this physics because of the large luminosity that can be
obtained and the fine resolutions of the spectrometers.
Furthermore, the continuous-wave electron beam significantly
enhances the signal to background ratio for coincidence
measurement. The advantage at JLab is that there is no need to
build new detector to search for the new particle
$Z(uuud\bar{s})$, since the existence or non-existence of
$Z(uuud\bar{s})$ can be re-constructed using the missing mass
technique by detecting the scattered electron and the final state
$K^-$ in coincidence. We do not need to know the explicit
properties, such as the life time, width, mass, spin, parity, and
decay modes {\it et al.} for detecting such particle.

\begin{figure}[htb]
\begin{center}
\leavevmode {\epsfysize=3.8cm \epsffile{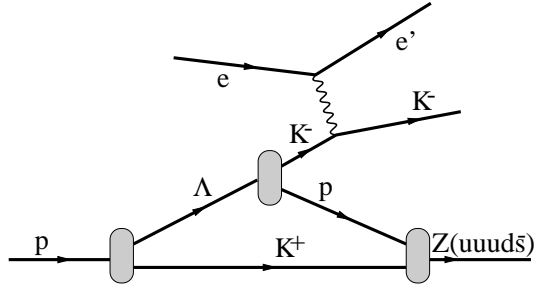}}
\end{center}
\caption[*]{\baselineskip 13pt The possible mechanism to produce
the exotic minimal pentaquark state $Z(uuud\bar{s})$ by the
electroproduction of $K^-$ from the proton target, i.e., the
physical process $p(e,e' K^-)Z(uuud\bar{s})$. } \label{gm1f1}
\end{figure}

We now estimate the probability of producing $Z(uuud\bar{s})$ from
the proton target by employing the baryon-meson fluctuation model
of the nucleon sea.\cite{Bro96} We check the probability of
finding the quark configuration of $uuud\bar{s}$ and $s\bar{u}$
inside the proton. From the discussion of the strangeness content
of the nucleon sea,\cite{Bro96} we know that the lowest
baryon-meson fluctuation of the strange quark-antiquark pairs is
$p(uuds\bar{s})=\Lambda(uds)K^+(u\bar{s})$ and the probability of
finding such fluctuation should be on the order 3\%. We also know
from the Gottfried sum rule violation\cite{Kum97} that the
dominant baryon-meson fluctuation of the proton is
$p(uudd\bar{d})=n(udd)\pi^+(u\bar{d})$ and the probability of
finding such fluctuation should be on the order 15\%.\cite{Bro96}
From the SU(3) symmetry of octet baryons, we can estimate the
probability of finding the baryon-meson fluctuation
$\Lambda(udsu\bar{u})=p(uud)K^-(s\bar{u})$ inside $\Lambda$ as
around 10\%. Thus the probability of finding the
baryon-meson-meson fluctuation $p(uudu\bar{u}s\bar{s})
=\Lambda(udsu\bar{u})K^+(u\bar{s})=p(uud)K^-(s\bar{u})K^+(u\bar{s})$
of the proton should be around $3\% \cdot 10\%=0.3\%$ from a
convolution picture, as shown in Fig.~\ref{gm1f1}. When the
$K^-(s\bar{u})$ is knocked out by the incident electron beam, the
left spectator should be of the quark configuration
$p(uud)K^+(u\bar{s})=uuud\bar{s}$, which is the same as that of
the pentaquark state $Z(uuud\bar{s})$. From above discussion, we
can estimate the cross section for the process
$p(e,e'K^-)Z(uuud\bar{s})$ by largest as around 0.1 times that of
the process $p(e,e'K^+)\Lambda$ by removing the kinematical
factors, provided with the assumption that the left spectator
keeps as a single state particle $Z(uuud\bar{s})$. In fact, the
real situation might be not so optimistic as the best case
estimated above.

Since the five quarks of $Z(uuud\bar{s})$ might be grouped into
clusters rather than confined inside a bag, we still have
difficulty predicting the mass of the minimal pentaquark state
$Z(uuud\bar{s})$, and it is likely below or slightly above the
threshold of two free particle state of $p$ and $K^+$, i.e.,
$M=1432$ MeV, if such particle can exist in the laboratory. If the
mass of $Z(uuud\bar{s})$ is above $M=1432$ MeV, such particle
should have a short life time and a broad width. Also there should
be large background contribution above the threshold of
$p(e,e'K^-)K^+p$ to the $p(e,e'K^-)Z(uuud\bar{s})$ process.
Therefore the resolutions required for the experiment on
$Z(uuud\bar{s})$ are high. In this situation, the new particle
$Z(uuud\bar{s})$ might also be possible to be produced and
measured in the resonance channel $K^{+} + p \to Z(uuud\bar{s})
\to K^{+} +p $ by re-constructing the $Z(uuud\bar{s})$ from the
detected final state $K^+$ and $p$, provided with good
resolutions. In case $\Lambda(1405)$ is a minimal pentaquark state
with configuration
$\Lambda(1405)=X(uuds\bar{u})=p(uud)K^-(s\bar{u})$ or
$\Lambda(1405)=X(udds\bar{d})=n(udd)\bar{K}^0(s\bar{d})$,\cite{l1405}
we may even consider the possibility that
$Z(uuud\bar{s})=p(uud)K^+(u\bar{s})$ is one of the partner
particles of $\Lambda(1405)$ and make a clear prediction of its
mass $M \approx 1405$ MeV. It would be relatively easy to be
measured from $p(e,e'K^-)Z(uuud\bar{s})$ process if the mass of
$Z(uuud\bar{s})$ is below $1432$ MeV. The possible channels for
the background contribution should be $p(e,e'K^-)\Delta^{++}$ and
$p(e,e'K^-)\pi^+ p $ which involve the Cabibbo suppressed weak
interaction below the threshold of the process $p(e,e'K^-)K^+ p $,
and such contributions are small. The re-construction of
$Z(uuud\bar{s})$ would be clean without significant background
contribution. Thus any evidence for or against the existence of
$Z(uuud\bar{s})$ might be also helpful in clarifying the situation
concerning $\Lambda(1405)$.

We can not exclude the possibility that only certain specific
pentaquark states, not all of them, can exist in nature. For
example, the existence of $Z(uudd\bar{s})$ is more possible than
that of $Z(uuud\bar{s})$ considering the fact that the stable
pentaquark states may carry less net charge.\footnote{However, the
proton with charge 1 is more stable than the neutron with charge
0.} We can extend the study in this direction by systematic
exploration of various minimal pentaquark states with
configurations pointed out above, such as to search for
$Z(uudd\bar{s})$ from the physical process $D(e,e'K^-p)
Z(uudd\bar{s})$ or $D(e,e'\Lambda) Z(uudd\bar{s})$ at JLab. It is
interesting to note a recent suggestion\cite{Zp} on the search for
such a $Z(uudd\bar{s})$ particle from the $pp \to n\Sigma^+ K^+$
reaction by analyzing the $n K^+$ invariant mass spectrum. We also
point out here that the search of the existence of the exotic new
particle $Z(uuud\bar{s})$ can be combined with the planned program
of the physical process of $p(e,e'K^-K^+)p$ at JLab. Thus the
search of new particles $Z(uuud\bar{s})$ and $Z(uudd\bar{s})$ at
JLab is not necessarily in conflict with conventional studies,
even though the chance of finding such particles might be small.
From another point of view, even the confirmation of the
non-existence of $Z(uuud\bar{s})$ and $Z(uudd\bar{s})$ can enrich
our understanding of the hadronic structure and the strong
interaction.

In summary, we proposed in this letter the idea of the possible
existence of exotic hadrons of minimal pentaquark $qqqq\bar{q}$
states such as $Z(uuud\bar{s})$. We found that the exotic hadrons
of pentaquark $qqqq\bar{q}$ states can not find place in the
conventional quark model spectroscopy of $qqq$-baryons if the
flavor of $\bar{q}$ is different from any of the other four
quarks, thus such minimal pentaquark states can be clearly
identified as a new class of particles if there is any evidence
for the existence. We also discussed the possible configurations
for such minimal pentaquark states and pointed out that
$Z(uuud\bar{s})$ might be identified. We suggested the physical
process $p(e,e'K^-)Z(uuud\bar{s})$, which can be studied at JLab,
as an ideal process to search for the existence or non-existence
of the exotic pentaquark hadron $Z(uuud\bar{s})$. The possibility
of a search for the $Z(uudd\bar{s})$ state from the physical
process $D(e,e'K^-p) Z(uudd\bar{s})$ or $D(e,e'\Lambda)
Z(uudd\bar{s})$ at JLab is also pointed out. Further theoretical
and experimental explorations are required to support or rule out
the existence of such exotic minimal pentaquark states which have
no place in the conventional $qqq$-baryon spectroscopy.

\section*{Acknowledgments}
This work is supported by National Natural
Science Foundation of China under Grant No.~19605006 and the U.S.
Department of Energy under contract number DE-FC02-94ER40818.


\section*{References}


\end{document}